\documentclass[12pt,titlepage]{article}
\usepackage{amsmath}
\usepackage[font=small,format=plain,labelfont=bf,up,textfont=it,up]{caption}
\usepackage{graphicx}

\begin{document}

\title{Equivalence between the planar Dirac oscillator and a spin-$1/2$
fermion embedded in a transverse homogeneous magnetic  field{\footnote {To appear in Revista Brasileira de Ensino de F{\'i}sica}}\\{\small (Equival{\^e}ncia entre o oscilador de Dirac planar e um f{\'e}rmion de spin $1/2$ imerso em um  campo magn{\'e}tico homog{\^e}neo
transverso)}}
\date{}
\author{Antonio S. de Castro\thanks{
E-mail: antonio.castro@unesp.br} \\
\\
Departamento de F\'{\i}sica e Qu\'{\i}mica, \\
Universidade Estadual Paulista \textquotedblleft J\'{u}lio de Mesquita
Filho\textquotedblright, \\
Guaratinguet\'{a}, SP, Brasil}
\maketitle

\begin{abstract}
It is shown that a spin-$1/2$ fermion coupled to the axially symmetric electromagnetic vector potential  has the same matrix structure as that one for 
the planar Dirac oscillator. In particular, the planar Dirac oscillator can 
be interpreted as a charged particle minimally coupled to a transverse 
homogeneous magnetic field.  \newline
\newline
\noindent Keywords: Dirac oscillator, relativistic planar motion, transverse
homogeneous magnetic field.\newline
\newline
\newline
\newline
\newline
{\small \noindent Mostra-se que um f{\'e}rmion de spin $1/2$ acoplado ao
potencial eletromagn{\'e}tico vetorial axialmente sim{\'e}trico tem a mesma
estrutura matricial que aquela do oscilador de Dirac planar. Em particular,
o oscilador de Dirac planar pode ser interpretado como uma particula
carregada minimamente acoplada a um campo magn{\'e}tico homog{\^e}neo
transverso.}  \newline
\newline
{\small \noindent Palavras-chave: oscilador de Dirac, movimento relativ{\'i}%
stico planar, campo magn{\'e}tico homog{\^e}neo transverso.} 
\end{abstract}

\section{Introduction}

The Dirac oscillator is an exactly solvable model consisting in a nonminimal
coupling prescription in the Dirac equation with the resulting equation
linear in both momentum and position operators \cite{mos}. Recently, much
interest has been generated on the planar Dirac oscillator. In particular,
has been investigated the bound-state spectrum and its degeneracy \cite{vil}-%
\cite{fab}, and applications to quantum optical phenomena \cite{ber1}-\cite%
{ber2}. Addition of a transverse uniform magnetic field has triggered
further investigations related to the Aharonov-Bohm \cite{fer}-\cite{ama1}
and Aharonov-Bohm-Coulomb effects \cite{oli}, coherent states \cite{oje},
optical models \cite{man1}-\cite{hou} and graphene \cite{ama1}, \cite{qui}-%
\cite{hat}. It should be mentioned that the authors of Refs. \cite{man1} and 
\cite{man2} have correctly recognized that the planar Dirac oscillator
immersed in a transverse homogeneous magnetic field can be mapped on a pure
planar Dirac oscillator.

The present work shows in a simple way the exact equivalence between the
planar Dirac oscillator and the problem of a charged particle minimally
coupled to a transverse magnetic field. Beyond a content interesting and
easy to deal with by graduate students in Physics, this result is of great
importance to help to clear up disagreements relating to the bound states
and its degeneracy, to assist the mapping of the planar Dirac oscillator
onto quantum optical models and graphene, and to assert the appropriate
chirality of the system needed to look into the critical magnetic field and
the possible chirality quantum phase transition relevant to applications in
quantum optical models and graphene.

\section{Dirac equation with an axially symmetric electromagnetic vector
potential}

In the Minkowski space-time, the behavior of a spin-1/2 fermion of mass $m$
and electric charge $q$ interacting with a stationary magnetic field is
governed by the Dirac equation 
\begin{equation}
i\frac{\partial \Psi }{\partial t}=H\Psi =\left[ \overrightarrow{\alpha }%
\cdot \left( \overrightarrow{p}-q\overrightarrow{A}\right) +\beta m\right]
\Psi ,  \label{D1}
\end{equation}%
with $\overrightarrow{p}=-i\overrightarrow{\nabla }$\  (in natural units $%
\hbar =c=1$). Here we have used the minimal coupling prescription 
\begin{equation}
\overrightarrow{p}\rightarrow \overrightarrow{p}-q\overrightarrow{A}.
\end{equation}%
The magnetic field is described by $\overrightarrow{B}=\overrightarrow{%
\nabla }\times \overrightarrow{A}$, and the matrices $\overrightarrow{\alpha 
}$ and $\beta $ can be represented as%
\begin{equation}
\overrightarrow{\alpha }=\left( 
\begin{array}{cc}
0 & \overrightarrow{\sigma } \\ 
\overrightarrow{\sigma } & 0%
\end{array}%
\right) ,\qquad \beta =\left( 
\begin{array}{cc}
I_{2\times 2} & 0 \\ 
0 & -I_{2\times 2}%
\end{array}%
\right) .
\end{equation}%
where $I_{2\times 2}$ is the $2\times 2$ unit matrix and $\overrightarrow{%
\sigma }=\left( \sigma _{1},\sigma _{2},\sigma _{3}\right) $. The spinor $%
\Psi $ has four components and the $2\times 2$ Pauli matrices obey the
fundamental relation%
\begin{equation}
\sigma _{i}\sigma _{j}=\delta _{ij}I_{2\times 2}+i\sum_{k=1}^{3}\varepsilon
_{ijk}\sigma _{k},
\end{equation}%
where $\delta _{ij}$ is the Kronecker delta and $\varepsilon _{ijk}$ is the
Levi-Civita symbol. In cylindrical coordinates $(\rho ,\varphi ,x_{3})$ one
has $\rho =|\overrightarrow{\rho }|=\sqrt{x_{1}^{2}+x_{2}^{2}}$ and $\varphi
=\arctan (x_{2}/x_{1})$ with coordinate unit vectors%
\begin{eqnarray}
\widehat{\rho } &=&\cos \varphi \,\widehat{e}_{1}+\sin \varphi \,\widehat{e}%
_{2}  \notag \\
\widehat{\varphi } &=&-\sin \varphi \,\widehat{e}_{1}+\cos \varphi \,%
\widehat{e}_{2} \\
\widehat{e}_{3} &=&\widehat{e}_{3},  \notag
\end{eqnarray}%
and%
\begin{equation}
\overrightarrow{\nabla }=\widehat{\rho }\,\frac{\partial }{\partial \rho }+%
\frac{\widehat{\varphi }}{\rho}\,\frac{\partial }{\partial \varphi }+\widehat{e}%
_{3}\,\frac{\partial }{\partial x_{3}}.
\end{equation}%
The axially symmetric electromagnetic vector potential 
\begin{equation}
\overrightarrow{A}=A_{\varphi }(\rho )\,\widehat{\varphi }\,  \label{pot}
\end{equation}%
gives a transverse magnetic field%
\begin{equation}
\overrightarrow{B}=B(\rho )\,\widehat{e}_{3},
\end{equation}%
with%
\begin{equation}
B(\rho )=\frac{1}{\rho }\frac{d\left[ \rho A_{\varphi }(\rho )\right] }{%
d\rho }.
\end{equation}%
Use of the axially symmetric electromagnetic vector potential allows the
Hamiltonian to be written as%
\begin{equation}
H=-i\alpha _{\rho }\frac{\partial }{\partial \rho }-i\alpha _{\varphi
}\left( \frac{1}{\rho }\frac{\partial }{\partial \varphi }-iqA_{\varphi
}\right) -i\alpha _{3}\frac{\partial }{\partial x_{3}}+\beta m,  \label{D3}
\end{equation}%
where $\alpha _{\rho }=\overrightarrow{\alpha }\cdot \widehat{\rho }$ and $%
\alpha _{\varphi }=\overrightarrow{\alpha }\cdot \widehat{\varphi }$, with%
\begin{equation}
\sigma _{\rho }=\overrightarrow{\sigma }\cdot \widehat{\rho }=\left( 
\begin{array}{cc}
0 & e^{-i\varphi } \\ 
e^{+i\varphi } & 0%
\end{array}%
\right) ,\qquad \sigma _{\varphi }=\overrightarrow{\sigma }\cdot \widehat{%
\varphi }=\left( 
\begin{array}{cc}
0 & -e^{-i\varphi } \\ 
e^{+i\varphi } & 0%
\end{array}%
\right) .
\end{equation}

\section{The exact equivalence with the planar Dirac oscillator}

It is remarkable that $A_{\varphi }(\rho )$ in the second term of the
Hamiltonian (multiplied by $\alpha _{\varphi }$) expressed by (\ref{D3}) can
be moved to the first term (multiplied by $\alpha _{\rho }$). This happens
because $\sigma _{\varphi }=i\sigma _{\rho }\sigma _{3}$ in such a way that $%
\alpha _{\varphi }=i\alpha _{\rho }\Sigma _{3}$. Here, 
\begin{equation}
\Sigma _{3}=\left( 
\begin{array}{cc}
\sigma _{3} & 0 \\ 
0 & \sigma _{3}%
\end{array}%
\right) .
\end{equation}%
Therefore, the Hamiltonian expressed by (\ref{D3}) can also be written as%
\begin{equation}
H=-i\alpha _{\rho }\left( \frac{\partial }{\partial \rho }+q\Sigma
_{3}A_{\varphi }\right) -i\alpha _{\varphi }\frac{1}{\rho }\frac{\partial }{%
\partial \varphi }-i\alpha _{3}\frac{\partial }{\partial x_{3}}+\beta m,
\end{equation}%
or equivalently,%
\begin{equation}
H=\overrightarrow{\alpha }\cdot \left( \overrightarrow{p}-iq\Sigma
_{3}A_{\varphi }\,\widehat{\rho }\right) +\beta m.  \label{D2}
\end{equation}%
This is an extraordinary result. The Hamiltonian expressed by (\ref{D2}) has
the same matrix structure as that one of the planar Dirac oscillator but
with a more general radial potential function due to the more general form for the axially symmetric electromagnetic vector potential $A_{\varphi }(\rho )$.
The problem of a charged particle minimally coupled to a transverse magnetic
field and the planar Dirac oscillator become indistinguishable when the
transverse magnetic field is uniform ($A_{\varphi }=B\rho /2$) and the
cyclotron frequency $|q|B/(2m)$ is identified with the frequency of the
Dirac oscillator.

\section{Final remarks}

For short, we showed that the planar Dirac oscillator for an electrically
charged particle can be interpreted as the problem describing a spin-$1/2$
fermion minimally coupled to a transverse homogeneous magnetic field. Hence,
their bound-state spectra and degeneracies are undoubtedly the same.
Applications of the planar Dirac oscillator to describe quantum optical
phenomena are also equivalent to applications of a transverse homogeneous
magnetic field. Addition of a transverse uniform magnetic field to a planar
Dirac oscillator clearly appears to be redundant.

\section*{Acknowledgement}

This work was supported in part by means of funds provided by CNPq (grant
304743/2015-1). 

\end{document}